\documentclass[twocolumn]{mn2e}
\usepackage{epsfig}
\usepackage{graphicx}

\def\Msun{\>{\rm M_{\odot}}}

\newcommand{\gtsim}{\mathrel{\hbox{\rlap{\lower.55ex \hbox {$\sim$}}
                   \kern-.3em \raise.4ex \hbox{$>$}}}}
\newcommand{\ltsim}{\mathrel{\hbox{\rlap{\lower.55ex \hbox {$\sim$}}
                   \kern-.3em \raise.4ex \hbox{$<$}}}}

\title[SMBH Growth Phase in Field AGNs]{Method for Determining AGN Accretion Phase in Field Galaxies}

\author[Micic, M., et al.]
{~Miroslav Micic$^1$\thanks{E-mail: micic@aob.rs},
~Nemanja Martinovi\'c$^1$,
~Manodeep Sinha$^2$ \\
$^1$ Astronomical Observatory Belgrade \\
$^2$ Department of Physics \& Astronomy, Vanderbilt University 
}

\begin{document}
\maketitle

\begin{abstract}

Recent observations of AGN activity in massive galaxies 
(log M$_*$/$\Msun$ $>$ 10.4) show that: 1) at z$<$1, 
AGN-hosting galaxies do not show enhanced merger signatures 
compared to normal galaxies, 2) also at z$<$1, most AGNs are 
hosted by quiescent galaxies; and 3) at z$>$1, percentage of 
AGNs in star forming galaxies increases and becomes comparable 
to AGN percentage in quiescent galaxies at z$\sim$2. How can 
major mergers explain AGN activity in massive quiescent galaxies 
which have no merger features and no star formation to indicate 
recent galaxy merger? By matching merger events in a cosmological 
N-body simulation to the observed AGN incidence probability in 
the COSMOS survey, we show that major merger triggered AGN activity 
is consistent with the observations. By distinguishing between 
``peak'' AGNs (recently merger triggered and hosted by star 
forming galaxies) and ``faded'' AGNs (merger triggered a long 
time ago and now residing in quiescent galaxies), we show that 
the AGN occupation fraction in star forming and quiescent galaxies 
simply follows the evolution of the galaxy merger rate. Since the 
galaxy merger rate drops dramatically at z$<$1, the only AGNs left 
to be observed are the ones triggered by old mergers and are now 
in the declining phase of their nuclear activity, hosted by quiescent 
galaxies. As we go toward higher redshifts, the galaxy merger rate 
increases and the percentages of ``peak'' AGNs and ``faded'' AGNs 
become comparable.

\end{abstract}


\section{INTRODUCTION}

Galaxies residing outside of galaxy clusters are known as 
field galaxies. Their name implies a certain level of isolation; 
either in time between major interactions ($\sim$ 3 Gyr, 
Verdes-Montenegro et al. 2005), or through the surrounding 
environmental density (Dressler 1980). 

The topic of this paper is AGN activity in field elliptical 
galaxies. These are massive (log M$_*$/$\Msun$ $>$ 10.4) 
galaxies, thought to be formed in gas rich major mergers of 
disk/spiral galaxies (Toomre 1977, White 1978, 1979, Gerhard 
1981, Negroponte $\&$ White 1983, Barnes 1988, Hernquist 1989, 
Barnes $\&$ Hernquist 1996, Naab, Jesseit $\&$ Burkert 2006, 
Novak et al. 2012).

We focus on field AGNs because AGNs in massive elliptical 
galaxies are a field phenomena. Hwang et al. 2012 have studied 
a sample of almost a million SDSS galaxies. They found factor 
of three larger AGN fraction in the field compared to clusters. 
At higher redshift this increase is even more pronounced. 
Martini, Sivakoff $\&$ Mulchaey 2009 found an increase by 
factor of 8 at redshift $z=1$. 

Galaxy mergers are also a field phenomena. Low velocity dispersion 
in galaxy groups in the field leads to ``slow encounters'' 
(Binney $\&$ Tremaine 1987) which are necessary for the merger 
to occur. ``Fast encounters'' are a characteristic of galaxy 
clusters. Energy input and dynamical friction scale as v$^{-2}$ 
(Binney $\&$ Tremaine 1987) and do not lead to the merger but 
rather small perturbations which can fuel a low luminosity 
AGN (Lake, Katz $\&$ Moore 1998).

For a long time, major galaxy mergers have been a main 
mechanism for driving AGN activity (Sanders et al. 1988, 
Barnes $\&$ Hernquist 1996, Cavaliere $\&$ Vittorini 2000, 
Menci et al. 2004, Croton et al. 2006, Hopkins et al. 2006, 
Menci et al. 2008), both supermassive black hole (SMBH) 
accretion and star formation (Sanchez et al. 2004, Bohm et al. 2007, 
Schawinski et al. 2007, Silverman et al. 2008, Rafferty et al. 2011, 
Hopkins 2012). Observational evidence indicates postmerger 
features in galaxies hosting AGNs and quasars (Surace $\&$ 
Sanders 1999, Surace, Sanders $\&$ Evans 2000, Canalizo $\&$ 
Stockton 2000, 2001) lending credence to the theoretical picture 
of mergers as drivers of AGN activity. Fiore et al. 2012 found 
that theoretical models using galaxy interactions as AGN 
triggering mechanism are able to reproduce the high redshift 
($z$=[3, 7]) AGN luminosity functions. The AGN fraction is 
higher in galaxy pairs (Silverman et al. 2011, Ellison et al. 2011, 2013).

This entire model was challenged recently (Gabor et al. 2009, 
Darg et al. 2010, Cisternas et al. 2011, Kocevski et al. 2012). 
Cisternas et al. 2011 found that 85 $\%$ of galaxies with AGNs 
do not show evidence of a previous merger at $z \leq$ 1, which 
is consistent with the merger fraction of non-active galaxies.

Schawinski et al. 2011, Kocevski et al. 2012, and Simmons et al. 2012 
showed that at z$\leq$ 3 there is a high disk fraction in AGN hosts. 
Bohm et al. 2013 found that morphologies of the AGN hosts are similar 
to undisturbed galaxies. These observations suggested that secular 
evolution is responsible for SMBH growth, at least at low $z$. These 
could be internal processes such as bar-driven gas inflow (Kormendy 
and Kennicutt 2004), and stellar wind (Ciotti and Ostriker 2007, Ciotti, 
Ostriker $\&$ Proga 2010, Cen 2012). At that moment it seemed that 
secular evolution is the dominant mechanism behind the activity of low 
luminosity AGNs, while major mergers of galaxies are responsible for 
luminous AGNs. 

Theoretical works also support this picture (Lapi et al. 2006, 
Hopkins, Kocevski $\&$ Bundy 2014). Hopkins, 
Kocevski $\&$ Bundy 2014 combined both merger and non-merger 
triggering of AGNs in semi-empirical model and found that 
secular (stochastic) fuelling is dominant in low luminosity 
AGNs which host SMBHs with mass $\leq 10^7 \Msun$. For luminous 
AGNs hosting SMBHs with masses $\geq 10^8 \Msun$ it accounts 
for just $\sim 10 \%$ of black hole (BH) growth. This is 
consistent with the observations of post-starburst quasars 
(PSQs) which show that PSQs with lower luminosities reside 
in disk/spiral galaxies, while more luminous PSQs reside in 
early type galaxies (Cales et al. 2013).

However, Villforth et al. 2014 analysed the morphological properties 
of AGN host galaxies as a function of AGN and host galaxy luminosity 
and compared them to a carefully matched sample of control galaxies 
in the redshift range $z$ = [0.5, 0.8] and luminosity range 
log L$_{\rm X}$ [erg/s] = [41, 44.5]. They found no increase in the 
prevalence of merger signatures with AGN luminosity and concluded that 
major mergers, even for higher luminosities, either play only a very 
minor role in the triggering of AGN or time delays are too long for 
merger features to remain visible. This conclusion questions galaxy 
mergers as drivers of any AGN activity.

In this paper we apply Shen 2009 SMBH growth model to the 
dark matter halo (DMH) merger trees in cosmological N-body 
simulation, in order to test if merger driven AGN activity 
is consistent with the activity of the observed AGNs in 
massive galaxies of COSMOS survey (Bongiorno et al. 2012). 

In the first part of the paper we determine initial BH mass, 
and final ({\bf true}) BH mass in the merger trees and then 
we use Shen 2009 AGN light curve model to grow initial BHs 
into final BHs. Next, we find our best fit model by matching 
it to the observed AGN luminosity function, active BHs mass 
function, duty cycle, and bias factor.

In the second part of the paper, we replace our best fit AGNs 
with COSMOS AGNs. We do this by using probability functions 
for galaxies to host AGNs in COSMOS, to determine probable 
AGN luminosities. We proceed with Monte Carlo procedure 
(40,000 realisations) where we replace peak luminosities in 
our best fit model with the COSMOS AGN luminosities. As the 
result, every postmerger halo has a 40,000 possible final 
BHs predicted by the model. Finally, we compare {\bf predicted} 
BH masses to the {\bf true} BH masses. 
We calculate the percentage of realisations when predicted 
SMBH mass is at least as large as the true SMBH mass. If that 
percentage is high, then the observed luminosity is the peak 
AGN luminosity. Otherwise, AGN is observed in the declining 
phase of its nuclear activity. This would place it in a massive, 
red, elliptical galaxy long after merger features can be detected, 
but its activity would still be consistent with merger driven model.

AGN hosts in COSMOS survey are mainly massive, red galaxies. 
Hence, their AGNs could potentially be merger driven, passed 
their peak activity during Green Valley, and in the declining 
Red Sequence phase. This interpretation would be consistent 
with the merger driven scenario for AGN activity and with the 
recent Schawinski et al. 2014 scheme for galaxy evolution.

We describe our method in section 2 and introduce two models 
based on initial BH mass function. In section 3, we present 
our best fit model. In section 4, we determine the phase
of AGNs activity is COSMOS survey. We discuss the implications 
of our results, in section 5.

\begin{figure*}
\vspace{10mm}
\epsfig{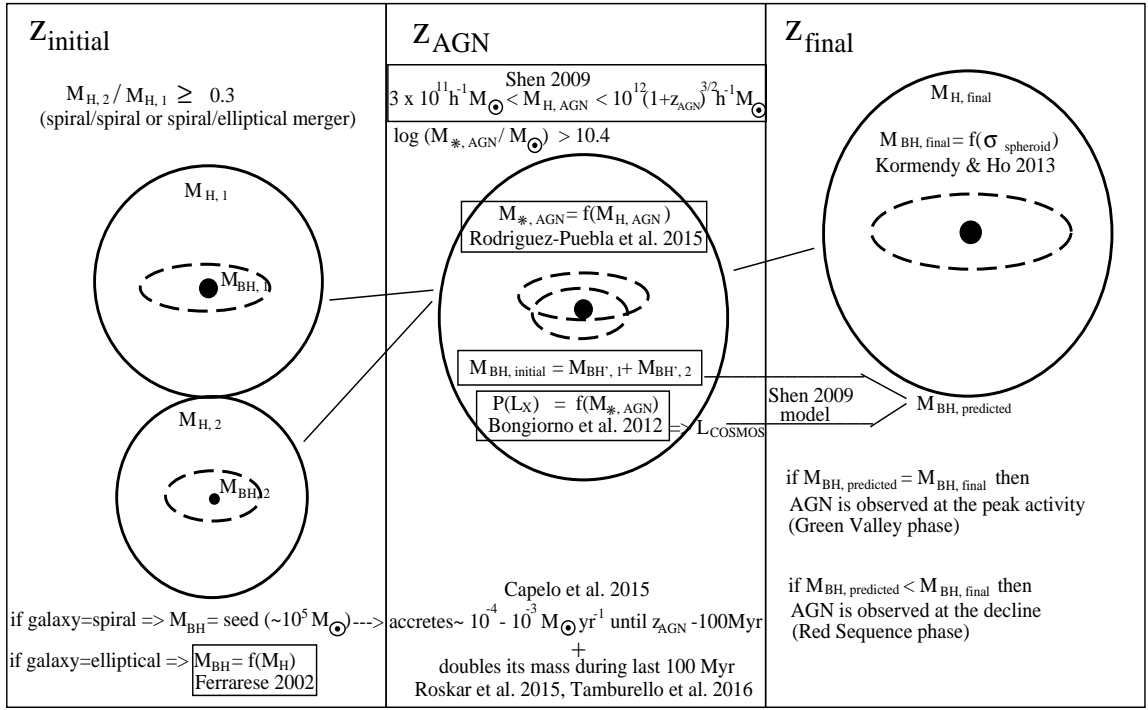}
\caption{Graphical sketch showing the main steps of the methodology.}
\end{figure*}

\section{METHOD}

The three major components in our model are: dark matter 
halo (DMH) merger trees from cosmological N-body simulation; 
Shen 2009 fit-by-observations semi-analytical model for 
major merger driven growth of SMBH; and Bongiorno et al. 2012
study of $\sim$ 1700 AGNs and their host galaxies in COSMOS 
field survey.

The main idea is to track ``field DMHs'' undergoing major 
mergers in N-body cosmological simulation. Use Shen 2009 
SMBH growth model and match it to the observations. Then 
we overlay our field with COSMOS field, match simulated 
galaxies to the observed COSMOS galaxies and assign COSMOS 
AGNs to them. Find if the observed AGNs are at their peak 
activity or in the declining phase.

Here is the outline of our model presented in figure 1.

\begin{enumerate}
\item z$_{\rm initial}$ is the redshift of DMH merger. 
Halos touch and the smaller halo is inside the larger 
halo at all later times. 

\item M$_{\rm H, 1}$ and M$_{\rm H, 2}$ are masses of 
merging halos at z$_{\rm initial}$. We consider major 
mergers only, when mass ratio of merging halos is $\geq$ 0.3. 

\item If merging halo did not have major merger in its 
history, we assume that halo hosts a spiral galaxy. If 
halo had a major merger before, we assume it hosts an 
elliptical galaxy. 

\item We seed spiral galaxies with pristine BHs 
($\sim$ 10$^5$ - 10$^6$ $\Msun$) and elliptical galaxies with BHs 
from Ferrarese 2002 M$_{\rm BH}$ - M$_{\rm DMH}$ relation.
Masses of BHs hosted by merging halos are M$_{\rm BH, 1}$ 
and M$_{\rm BH, 2}$. Mergers of two 
elliptical galaxies do not trigger AGN activity 
(dry mergers). 

\item $z_{\rm AGN}$ is the redshift when smaller halo can not be 
identified anymore inside the larger one which means that the 
merger of DMHs has finished. We assume that mergers of their galaxies 
and black holes have finished too, and that accretion onto the 
new SMBH starts and enters AGN phase which has pre-peak and peak activity.

\item Even before central BHs (M$_{\rm BH, 1}$ and M$_{\rm BH, 2}$) 
form binary (BHB), they accrete at $\sim$ 10$^{-3}$ - 10$^{-4}$ $\Msun$ $\rm yr^{-1}$ 
rate (Capelo et al. 2015), from z$_{\rm initial}$ at R$_{\rm vir}$ 
separation until BHB forms at $\sim$ kpc distance. This is the 
``pre-BHB accretion'' phase. During the last $\sim$ 100 Myr before 
BHs merge (at z$_{\rm AGN}$), binary overcomes last couple of kpc 
and accretion increases to double the BH mass (Roskar et al. 2015, 
Tamburello et al. 2016). This is the ``BHB accretion'' phase.
New masses of central BHs after these two accretion episodes are 
M$_{\rm BH', 1}$ and M$_{\rm BH', 2}$.

\item SMBH mass entering AGN phase at z$_{\rm AGN}$, is then 
M$_{\rm BH,initial}$ = M$_{\rm BH',1}$ + M$_{\rm BH',2}$.
 
\item M$_{\rm H, AGN}$ is the mass of DMH hosting an AGN at 
z$_{\rm AGN}$. We adopt Shen 2009 model for AGN activity in 
field galaxies. This model implies halo mass range 
3 $\times$ 10$^{11}$ h$^{-1}\Msun$ $<$ M$_{\rm H, AGN}$ $<$ 10$^{12}$ (1+z$_{\rm AGN})^{3/2}$h$^{-1}\Msun$.

\item M$_{\rm *,AGN}$ is the mass of the galaxy hosting an 
AGN at z$_{\rm AGN}$. We consider only AGNs in massive galaxies 
(log(M$_{*,\rm AGN}$/$\Msun$) $>$ 10.4). Galaxy mass is obtained 
by using Rodriguez-Puebla et al. 2015 M$_*$ - M$_{\rm DMH}$
relation.

\item P(L$_{\rm X}$) is the probability of a galaxy to host 
an AGN of a given luminosity at z$_{\rm AGN}$ 
(Bongiorno et al. 2012). From it we obtain L$_{\rm X}$ and 
calculate bolometric AGN luminosity L$_{\rm COSMOS}$.

\item M$_{\rm BH, predicted}$ is the SMBH mass predicted by 
Shen 2009 model, when M$_{\rm BH,initial}$ is the input 
parameter given in point (vii) and the peak luminosity is replaced by L$_{\rm COSMOS}$.

\item z$_{\rm final}$ is the redshift of the postmerger 
halo M$_{\rm H, final}$.

\item M$_{\rm BH, final}$ is the ``true'' mass of the postmerger 
BH, derived from M$_{\rm BH}$ - $\sigma_{\rm sph}$ relation, 
calibrated to the local M$_{\rm BH}$ - M$_{\rm DMH}$ relation 
(Ferrarese 2002). 

\item If the observed COSMOS AGN is at the peak activity, 
then M$_{\rm BH, predicted}$ has to be at least as large 
as M$_{\rm BH, final}$. Otherwise, AGN is in the declining 
phase of its nuclear activity.

\end{enumerate}

In the following sections we describe simulation, 
data, and modelling in more details.

\subsection{Cosmological N-body Simulation}

Using GADGET2 (Springel, Yoshida $\&$ White 2001, Springel et al. 2005), 
we performed a high-resolution cosmological N-body simulation within a 
comoving periodic box with size of 130 Mpc$^3 $. WMAP5-like 
(Komatsu et al. 2009) cosmology was used ($\Omega_{\rm M}=0.25$, 
$\Omega_{\Lambda}=0.75$, $n_s=1$, $\sigma_8$=0.8 and h=0.7) from 
$z=599$ to $ z=0$ (84 snapshots). Initial conditions were computed with 
the 2LPT code (Crocce, Pueblas $\&$ Scoccimarro 2006). Simulation 
utilises 512$^3$ dark matter particles for a mass resolution of 
1.14 x 10$^9$ $\Msun$.

We generated halo catalogues using ROCKSTAR (Behroozi, 
Wechsler $\&$ Wu 2013). ROCKSTAR combines friends-of-friends 
(FOF), phase-space and spherical overdensity analysis in 
locating halos. Please see Behroozi, Wechsler $\&$ Wu 2013 
for details on the ROCKSTAR algorithm. The merger tree was 
generated using Consistent Merger Tree (Behroozi et al. 2013), 
a software package that is complementary with the ROCKSTAR 
halo finder.

\subsection{AGNs and galaxies in COSMOS survey}

Bongiorno et al. (2012) have studied $\sim$ 1700 AGNs in COSMOS 
field obtained by combining X-ray and optical spectroscopic 
selections. This is a highly homogeneous and representative 
sample of obscured and unobscured AGNs over a wide redshift 
range (0 $<$ z $<$ 4). By using Spectral Energy Distribution 
(SED) fitting procedure they have managed to separate host 
galaxies properties including the total stellar mass of galaxies 
hosting AGNs. One of their results is the probability of a 
galaxy to host an AGN of a given luminosity (P(L$_{\rm X}$)) 
as a function of stellar mass in three redshift bins: 
[0.3 - 0.8], [0.8 - 1.5], and [1.5 - 2.5] (Figure 14 in their 
paper, from here on F14). They grouped AGNs in four logarithmic 
X-ray (2 - 10 KeV) luminosity bins: [42.8 - 43.5], [43.5 - 44.0], 
[44.0 - 44.5], and [44.5 - 46.0] in erg/s units. They showed that for a fixed 
mass range, observed field galaxies are more likely to host 
less luminous AGNs. The probability that a field galaxy hosts 
an AGN decreases with increasing AGN luminosity.

\subsection{SMBH growth model}

We adopt Shen 2009 model for the hierarchical growth and 
evolution of SMBHs assuming that AGN activity is triggered 
in major mergers. This model uses a general form of light 
curve where BH first grows exponentially at constant luminosity 
Eddington ratio of $\lambda_0$=3 (Salpeter 1964) to 
L$_{\rm peak}$ at t = t$_{\rm peak}$, and then the luminosity 
decays monotonically as a power-law (Yu $\&$ Lu 2008).

Shen 2009 uses a variety of observations. Model adopts 
Hopkins et al. 2007 compiled AGN bolometric luminosity 
function data for both unobscured and obscured SMBH growth, 
It also incorporates quasar clustering observations and the 
observed Eddington ratio distributions.

Model successfully reproduces the observed AGN luminosity 
function and both the observed redshift evolution and 
luminosity dependence of the linear bias of AGN clustering.

The input parameters for the Shen 2009 model are 
M$_{\rm BH,initial}$ (mass of the BH entering AGN phase), 
L$_{\rm peak}$ (peak bolometric AGN luminosity), and
M$_{\rm BH, relic}$ (BH mass in the postmerger halo 
M$_{\rm DMH, post}$ immediately after the AGN phase). 
To match our nomenclature, we have renamed 
M$_{\rm BH, relic}$ to M$_{\rm BH, final}$.

In our model, values for the first parameter come from the 
numerical simulation combined with the semi-analytical 
modelling (details in section 2.4).

We calculate M$_{\rm BH, final}$ (details in section 2.6) 
from M$_{\rm BH}$ - $\sigma_{\rm sph}$ relation 
(Kormendy $\&$ Ho 2013) where $\sigma_{\rm sph}$ is the 
velocity dispersion of the stellar spheroid. $\sigma_{\rm sph}$ 
is correlated with V$_{\rm vir}$ by a constant (Ferrarese 2002). 
We set this constant to a value which reproduces z = 0 
Ferrarese 2002 M$_{\rm BH}$ - M$_{\rm DMH}$ relation. 
Since M$_{\rm BH}$ - $\sigma$ relation is expected to be 
non-evolving, one can find M$_{\rm BH}$ in M$_{\rm DMH, post}$ 
at any redshift. The outcome of this procedure is that the 
BH mass in high redshift halos, right after AGN phase, is 
overestimated. This is expected to occur as BH grows first, 
followed by postmerger halo growth through minor mergers and 
diffuse matter accretion. As we go toward lower redshifts, 
DMH growth catches up to SMBH growth to reproduce local 
Ferrarese relation (Figure 6). Hence, we consider
M$_{\rm BH, final}$ to be the ``true'' final BH mass.

L$_{\rm peak}$ is the peak bolometric luminosity 
(details in section 2.7) in the Shen 2009 light curve model, 
set to the value which guarantees that the accretion onto 
M$_{\rm BH,initial}$ produces M$_{\rm BH, final}$.
This is our (L$_{\rm peak, true}$) best fit model which 
reproduces the observed AGN activity, luminosity function, 
duty cycle and bias factor.  

After we demonstrate that our merger driven model 
reproduces observed AGN statistics, we test if the 
observed AGN activity in COSMOS survey corresponds to
the peak or to the declining activity. Now, instead 
of L$_{\rm peak, true}$, values for the peak luminosity 
(L$_{\rm COSMOS}$) are retrieved from the probability 
for a galaxy to host an AGN of a given luminosity 
(P(L$_{\rm X}$)) in COSMOS survey (details in section 
2.8). We use this probability to seed galaxies with 
AGNs in 40,000 Monte Carlo realisations and grow 
SMBHs according to Shen 2009 model (details in section 
2.9). The result is the probability that SMBHs grown 
in COSMOS AGNs match the true SMBHs grown in our 
best fit model.

\subsection{Halos, galaxies, black holes: Initial values}

We start by identifying major merger events in the merger 
trees of our cosmological N-body simulation. We define masses 
of merging halos as M$_{\rm H, 1}$ and M$_{\rm H, 2}$ at the 
time of the merger z$_{\rm initial}$ (figure 1). We also check 
if merging halos had major mergers previously. DMH without 
previous major merger is an ancient halo hosting disk/spiral 
galaxy with a large cold gas reservoir and the central BH 
that most likely formed through direct collapse of a gas cloud 
(Bromm $\&$ Loeb 2003, Begelman, Volonteri $\&$ Rees 2006, 
Begelman, Rossi $\&$ Armitage 2008). Latest observations 
(Mortlock et al. 2011) showed that BH seeds had to be massive 
($\sim$ 10$^5$ - 10$^6$$\Msun$) in order to grow $\sim$ 
10$^9\Msun$ BHs at redshift z $\sim$ 7.

The initial mass function (IMF) and the mass range of the 
seed BHs are unknown. These BHs settle at the centres of 
disk/spiral galaxies but their masses do not correlate with 
any of the galaxy properties. 

Growth of these initial BHs through accretion can occur even 
before they form a binary (BHB), during the early stages of 
the galaxy merger as galaxies go through subsequent pericentric 
passages (Capelo et al. 2015). As the major merger of galaxies 
proceeds, gravitational torques generate large-scale gas 
inflows that drive the gas down to sub-pc scale where it can
be accreted by the BH. Hence, BHs grow in a modest amount 
even before they form a binary (pre-BHB 
accretion). Modelling of this growth is a subject of numerous 
numerical studies. However, limited resolution and disparate
subgrid physics recipes led to a very different estimates of 
the BH accretion rates. Latest results (Hayward et al. 2014, 
Capelo et al. 2015) show that BH accretes at the rate 
10$^{-4}\Msun {\rm yr^{-1}}$ - 10$^{-3}\Msun {\rm yr^{-1}}$ 
for $\sim$ 1 Gyr before AGN phase.

After BH binary forms, accretion increases as BHs sink to overcome
the last couple of kiloparsecs between them. During these last
$\sim$ 100 Myr before BH merger, BHs double their masses
(Roskar et al. 2015, Tamburello et al. 2016). Assuming that 
Salpeter time is $\sim$50 Myr, corresponding Eddington ratio
is 0.35. This would mean that 10$^6\Msun$ BH accretes at rate 
of 0.1 $\Msun {\rm yr^{-1}}$, while 10$^7\Msun$ BH accretes at 
the rate of 1 $\Msun {\rm yr^{-1}}$ (over 100 Myr). 
This ``BHB accretion'' phase finishes with BH binary coalescence 
into a new BH which enters an AGN phase.

Unknown IMF for BH seeds, and rate of ``pre-BHB 
accretion'' are the main sources of uncertainty in our 
modelling. We overcome this issue by considering two models 
with the idea of constraining lower and upper end of 
possible initial BH mass. Our lower constraint model (M1) 
assumes log-normal IMF for BH seeds in the interval 
log (M$_{\rm BH}/\Msun$) = [4.5, 5.5] centered at 10$^5\Msun$, 
and pre-BHB accretion with a rate of 
$\dot{\rm m}$ = 10$^{-4}\Msun {\rm yr^{-1}}$ from 
z$_{\rm initial}$ until z$_{\rm AGN}$ minus 100 Myr. For a higher 
constraint model M2 we set log (M$_{\rm BH}/\Msun$) = [5.0, 6.0] 
centred at 10$^{5.5}\Msun$ and $\dot{\rm m}$ = 10$^{-3}\Msun {\rm yr^{-1}}$.
We seed DMHs with BHs by randomly choosing BH masses from 
these IMFs in Monte Carlo realisations. In the last 100 Myr before 
z$_{\rm AGN}$, we double the BH mass (BHB accretion).

If DMH already had a major merger in its history, we assume 
it hosts an elliptical galaxy. BH at the centre of an elliptical 
galaxy scales with the properties of the stellar spheroid but
also with the mass of the host DMH. We use Ferrarese 2002 
M$_{\rm BH}$ - M$_{\rm DMH}$ relation with $\pm$10$\%$ scatter 
to seed elliptical galaxies with central BHs. Mergers of two 
elliptical galaxies do not trigger AGN activity (dry mergers). 
If an elliptical galaxy merges with a spiral, there is no 
pre-BHB or BHB accretion onto the BH at the centre of the 
elliptical galaxy.

Initial mass of the BHs in both halos (M$_{\rm BH, 1}$ and 
M$_{\rm BH, 2}$) combined with pre-BHB and BHB accretion 
(if galaxy is spiral) produces M$_{\rm BH', 1}$ and M$_{\rm BH', 2}$.
Initial BH mass that enters AGN phase is then: 
M$_{\rm BH,initial}$ = M$_{\rm BH',1}$ + M$_{\rm BH',2}$.

\subsection{AGN phase}

At $z_{\rm AGN}$, initial BHs merge, form new BH (M$_{\rm BH,initial}$). 
Mass of the DMH hosting an AGN is then M$_{\rm H, AGN}$. Accretion 
onto M$_{\rm BH,initial}$ starts first with the pre-peak phase at 
super Eddington rate ($\lambda$ = 3) followed by the declining phase 
best described by Figure 2 in Shen 2009. We assume that AGN reaches 
its peak activity at $z_{\rm AGN}$. 

Note that in Shen 2009 model, AGN activity starts at the time when halos
merge (not galaxies). Hence, the AGN activity in their model is pushed
toward slightly higher redshifts. We find that the typical delay between
halo merger and consecutive galaxy merger is $\Delta$z = 0.2 in redshift 
space and it does not impact overall results.

We adopt Shen 2009 model for AGN activity in field galaxies. 
We consider halos in mass range  
3 $\times$ 10$^{11}$ h$^{-1}\Msun$ $<$ M$_{\rm H, AGN}$ $<$ 10$^{12}$ (1+z$_{\rm AGN})^{3/2}$h$^{-1}\Msun$.
If halo mass is too small mass, AGN activity can not be triggered, while 
overly massive halos can not cool gas efficiently and BH growth 
halts (especially at low redshift) (Shen 2009). This excludes 
high density environments (e.g galaxy clusters) from our model 
and we are left with the AGN activity in the field. We do find
that increasing the upper limit on host halo mass overpredicts the
AGN luminosity functions at low redshift (z $\leq$ 1).

Mass of the galaxy hosting an AGN is M$_{\rm *,AGN}$. Since 
the topic of this paper is to examine merger driven AGN 
activity in massive galaxies in the field, we consider 
galaxies with log(M$_{*,\rm AGN}$/$\Msun$) $>$ 10.4. In lower 
mass galaxies, SMBHs are more likely to accrete through 
secular processes related to channeling of the gas through 
bars or disk instabilities. 
 
Galaxy mass is obtained by using Rodriguez-Puebla et al. 2015 
M$_*$ - M$_{\rm DMH}$ relation for early type (elliptical) galaxies 
(equations 17 and 18 and Figure 5 in their paper) with scatter 
$\sigma_{\rm r}$ = $\pm$0.136 dex (equation 37 in 
Rodriguez-Puebla et al. 2015). Scatter determines galaxy mass 
in every Monte Carlo realisation.

\subsection{Halos, galaxies, black holes: Final values}

M$_{\rm H, final}$ is the mass of the postmerger halo 
(immediately after the AGN phase) hosting final (relic) 
SMBH. We chose to define the time z$_{\rm final}$ to be 
$\sim$ 100 Myr after AGN phase z$_{\rm AGN}$, located 
in the first consecutive snapshot. However, mass of 
the postmerger halo changes insignificantly in more than 
one snapshot after z$_{\rm AGN}$. In fact, our results
do not change even when we use M$_{\rm H, AGN}$ instead 
of M$_{\rm H, final}$. This occurs because at the time 
of galaxy merger, new halo has already formed and for 
some time after the AGN phase it stays intact. Later, 
it continues growing by minor mergers and diffuse matter 
accretion. This implies that at first, mass of the final 
(relic) SMBH (M$_{\rm BH, final}$) hosted by M$_{\rm H, final}$ 
will be overestimated when compared to the local Ferrarese 
M$_{\rm BH}$ - M$_{\rm DMH}$ relation. As M$_{\rm H, final}$ 
grows in mass over time, M$_{\rm BH, final}$ - M$_{\rm DMH, final}$ 
relation approaches Ferrarese relation. 
   
Since M$_{\rm BH}$ - $\sigma_{\rm sph}$ relation is expected 
to be non-evolving (Gaskell 2009, Shankar, Bernardi $\&$ Haiman 2009, 
Salviander, Shields $\&$ Bonning 2015, Shen et al. 2015), one can find M$_{\rm BH, final}$ in 
M$_{\rm DMH, final}$ at any redshift. First, one can rewrite 
equation (3) in Ferrarese 2002 as:

\begin{equation}
\frac{\rm V_{{\rm vir}}}{\rm 200 km s^{-1}} = (\frac{\rm M_{H, final}}{2.7 \times 10^{12} \Msun})^{1/3} ,
\end{equation}

Next, $\sigma_{\rm sph}$ = C $\times$ V$_{\rm vir}$. 
And from equation (7) in Kormendy $\&$ Ho 2013:

\begin{equation}
\frac{\rm M_{{\rm BH, final}}}{\rm 10^9 \Msun} = 0.309 (\frac{\rm \sigma_{sph}}{200 {\rm km s^{-1}}})^{4.38} ,
\end{equation}
with scatter $\sigma$ = $\pm$0.28 dex.

We find that for C = 0.77, our M$_{\rm BH, final}$ - M$_{\rm DMH, final}$ relation
at z=0 matches local Ferrarese relation. As we go toward higher redshifts,
Ferrarese relation evolves (figure 6) and M$_{\rm BH, final}$ is overestimated 
while M$_{\rm BH}$ - $\sigma_{\rm sph}$ does not evolve.

\subsection{Finding Best Fit Model}

Now that we have obtained M$_{\rm BH, initial}$ and M$_{\rm BH, final}$,
we can calculate L$_{\rm peak}$ necessary to produce M$_{\rm BH, final}$.
As already mentioned in section 2.5, we use Shen 2009 AGN light curve
with pre-peak exponential growth phase followed by post peak power-law decline.
To calculate L$_{\rm peak}$ we rewrite equation (29) in Shen 2009:

\begin{equation}
{\rm L_{peak}} = 3{\rm M_{BH, final}}{l_{\rm Edd}} (1-\frac{2{\rm lnf}}{3})^{-1}  , 
\end{equation}
and
\begin{equation}
f = \frac{3 l_{\rm Edd} {\rm M_{BH, initial}}}{{\rm L_{peak}}}
\end{equation}
where $l_{\rm Edd}$=1.26 $\times$ 10$^{38}{\rm ergs}^{-1}\Msun^{-1}$. 

The descending phase is presented by equation 24 in Shen 2009:

\begin{equation}
{\rm L(L_{peak}, t)} = {\rm L_{peak}} (\frac{\rm t}{\rm t_{peak}})^{-\rm \alpha}  , 
\end{equation}
where $\alpha$ = 2.5. Luminosities of all AGNs in all galaxies and
at all redshifts decrease three orders of magnitude from their peak 
luminosity in $\sim$ 2 Gyr.

We use M$_{\rm BH, initial}$, M$_{\rm BH, final}$, 
and L$_{\rm peak}$ to calculate AGN luminosity function, 
active SMBH mass function, duty cycle, and bias factor. 
We compare these to the observed values. We find that
both M1 and M2 models reproduce observations without 
any additional modelling or parameter fixing. We continue 
with M1 and M2 as our best fit models and later replace 
L$_{\rm peak}$ with L$_{\rm COSMOS}$ to find the AGN 
activity phase in COSMOS survey.

\subsection{Matching COSMOS AGNs to M$_{\rm *,AGN}$}

For $\sim$ 1700 AGNs in COSMOS field Bongiorno et al. 2012 presented
probability of a galaxy of a certain mass to host an AGN of a given 
luminosity as a function of stellar mass in three redshift bins: 
[0.3 - 0.8], [0.8 - 1.5], and [1.5 - 2.5] (F14). They group AGNs in 
four X-ray (2 - 10 KeV) luminosity bins in logarithm space: 
[42.8 - 43.5], [43.5 - 44.0], [44.0 - 44.5], and [44.5 - 46.0] (erg/s units). 
Masses of their AGN hosting galaxies are also separated in logarithmic bins:
[9.0, 10.0], [10.0, 10.4], [10.4, 10.7], [10.7, 10.9], [10.9, 11.2] ($\Msun$ units).   

How to pick a luminosity from F14 and assign it to our M$_{\rm BH,initial}$?
We do this by grouping our simulated galaxies at the moment their 
M$_{\rm BH,initial}$ should start accreting.

We determine z$_{\rm AGN}$, M$_{\rm BH,initial}$, and M$_{\rm *, AGN}$ 
for every merger in our simulation and group them into redshift bins:
\begin{equation}
\Delta\rm z = [0.3 - 0.8; 0.8 - 1.5; 1.5 - 2.5],
\end{equation}
and galaxy log-mass bins:
\begin{equation}
\Delta\rm M_{\rm *} = [10.4 - 10.7; 10.7 - 10.9; 10.9 - 11.2],
\end{equation}
since we study AGNs in massive galaxies only.

\begin{figure*}
\vspace{10mm}
\epsfig{figure=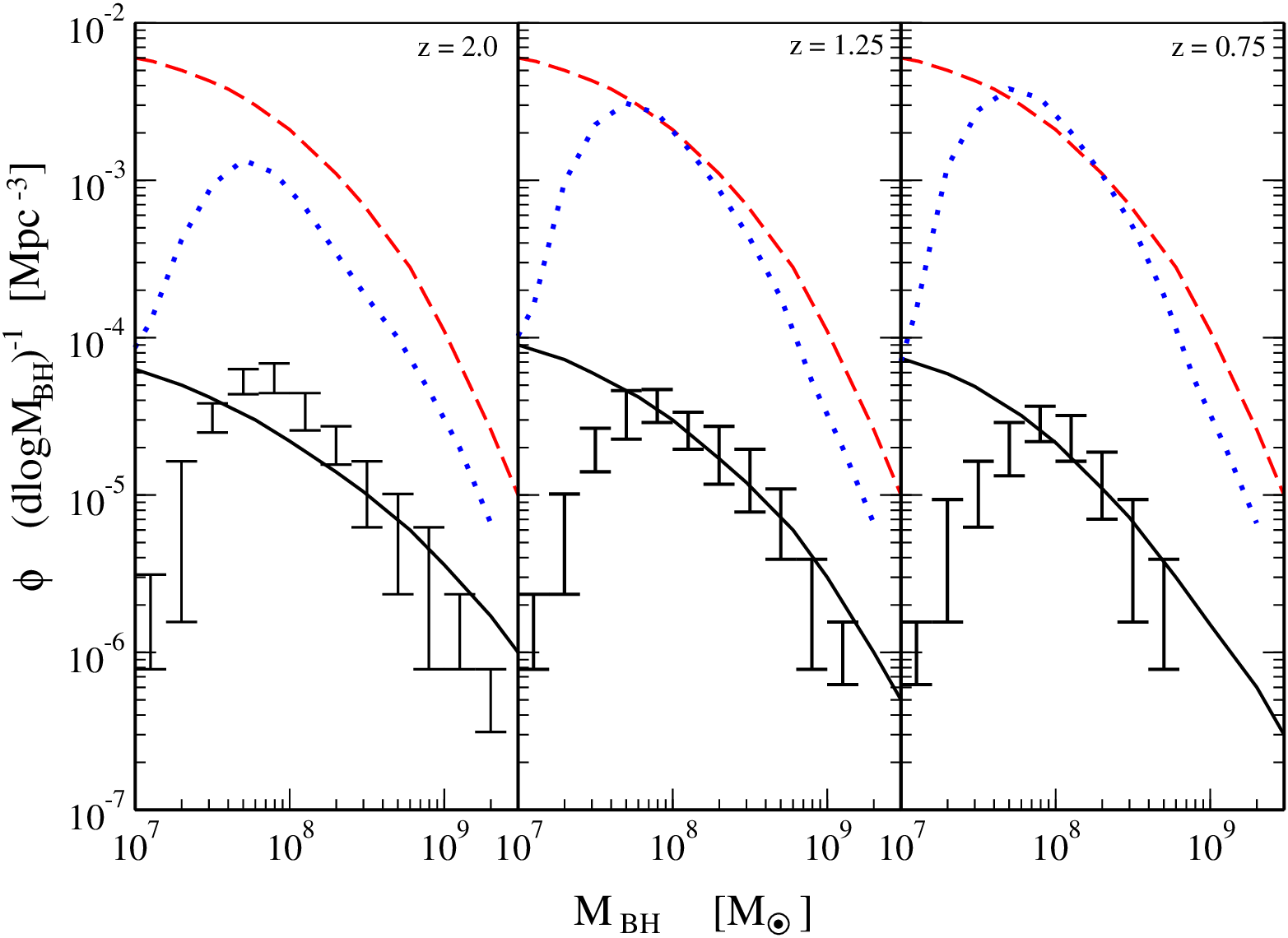, height=0.5900\textwidth, width=0.950\textwidth}
\caption{Black hole mass function at three redshifts z = [2.00, 1.25, 0.75], 
for active black holes only, in AGNs with log L$_{\rm X}$ $\rm {[erg/s]}$ $\geq$ 43, 
where X = [2 - 10] KeV. Horizontal and vertical bars show the full range of 
masses in our Monte Carlo realisations. Dotted, blue line shows our BH mass function
for all BHs. Overplotted as thick black line 
is active BH mass function for the same luminosity range from observations (HELLAS2XMM)
of La Franca et al. 2005 (presented in Fiore et al. 2012). Also, in dashed red
line is local BH mass function for all black holes (Merloni $\&$ Heinz 2008). 
Our best fit model is a good match to the observations. } 
\end{figure*}

In this manner we obtain nine $\Delta\rm z$-$\Delta\rm M_{*}$ intervals.
The number of galaxies belonging to each $\Delta\rm z$-$\Delta\rm M_{*}$ interval
is N$_{\rm *, AGN}$. Next we match these $\Delta\rm z$-$\Delta\rm M_{*}$ 
intervals to the $\Delta\rm z$-$\Delta\rm M_{*}$ intervals in F14. 
According to F14, galaxies can host AGNs with luminosities in intervals:

\begin{equation}
\Delta\rm L_{\rm X} = [42.8 - 43.5; 43.5 - 44.0; 44.0 - 44.5; 44.5 - 46.0].
\end{equation}

So the BHs in N$_{\rm *, AGN}$ simulated galaxies can be 
assigned with any of the luminosities from $\Delta$L$_{\rm X}$ intervals. 
How these luminosities should be assigned is determined by the 
probability P$_{\rm AGN,i}$ (data points in F14) defined for 
every $\Delta$L$_{\rm X, i}$. 

P$_{\rm AGN,i}$ in F14 tells us that every galaxy in a specific 
$\Delta\rm z$-$\Delta\rm M_{*}$ interval is more likely to host
low luminosity AGN. 

Since the number of galaxies in each $\Delta\rm z$-$\Delta\rm M_{*}$ 
interval is N$_{\rm *, AGN}$, then the number of times a luminosity 
should be drawn from each luminosity interval $\Delta$L$_{\rm X, i}$ is:

\begin{equation}
N_{\rm L, i} = \frac{P_{\rm AGN,i}}{\Sigma P_{\rm AGN,i}} \times N_{\rm *, AGN}  , 
\end{equation}

Largest N$_{\rm L,i}$ is for the interval $\Delta$L$_{\rm X, i}$ with 
smallest luminosities. The sum of N$_{\rm L,i}$ is equal to N$_{\rm *, AGN}$.

Next we randomly draw luminosities N$_{\rm L,i}$ times from every 
corresponding $\Delta$L$_{\rm X, i}$ and we randomly assign them to 
N$_{\rm *, AGN}$ galaxies.

\begin{figure*}
\vspace{10mm}
\epsfig{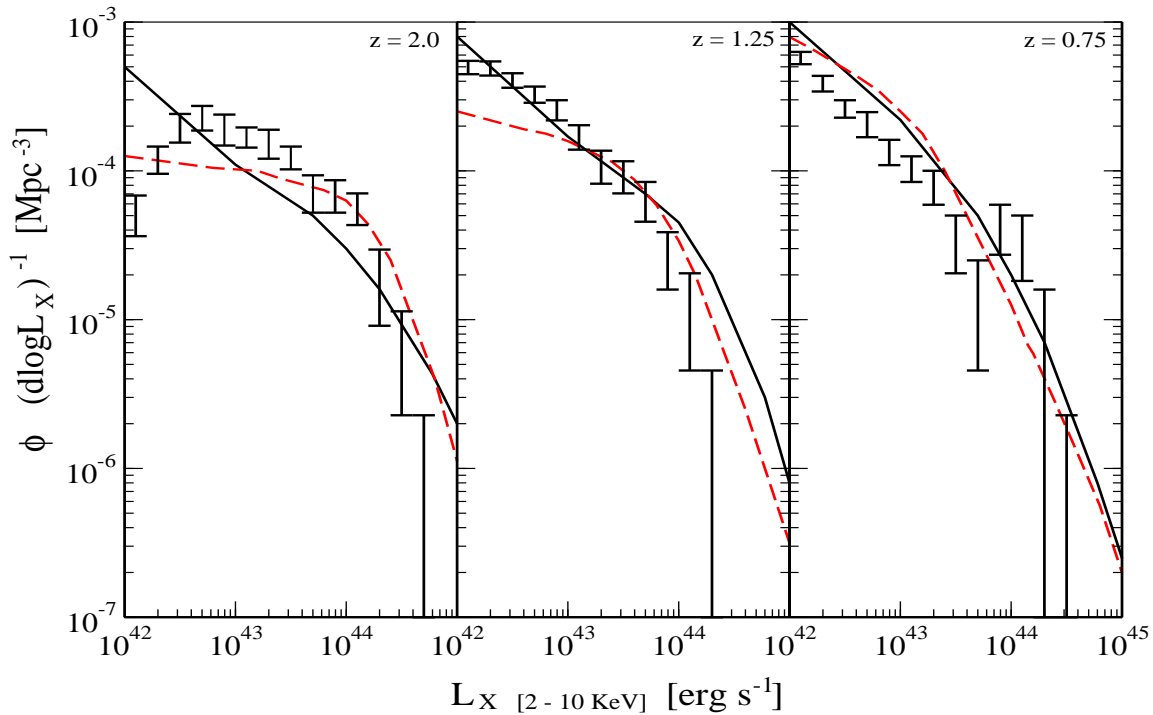}
\caption{AGN luminosity function at three redshifts z = [2.00, 1.25, 0.75]. 
Horizontal and vertical bars present the full range in our best fit model. 
Overplotted as thick black line is AGN luminosity function from observations 
(HELLAS2XMM) of La Franca et al. 2005 (presented in Fiore et al. 2012). 
In dashed red line is AGN luminosity function from a large combination of 
X-ray surveys including XMM and Chandra COSMOS survey (Miyaji et al. 2015). 
Our best fit model is a good match to the observations although we slightly 
underpredict luminosity function toward lower redshifts.} 
\end{figure*}

This is the first out of 40,000 Monte Carlo realisations where we 
draw luminosity values to be assigned to AGNs in each 
$\Delta\rm z$-$\Delta\rm M_{*}$ interval. Thus, for each 
M$_{\rm *, AGN}$ we have a set of 40,000 COSMOS AGN luminosities.
Since these are X-ray luminosities, we use equation 2) in 
Hopkins et al. 2007 to calculate bolometric luminosities. We address 
these luminosities as L$_{\rm COSMOS}$. Next, we replace L$_{\rm peak}$ 
in our best fit model with L$_{\rm COSMOS}$.

\subsection{Modelling SMBH growth in COSMOS}

With calculated M$_{\rm BH,initial}$ (mass of the BH entering 
AGN phase) and L$_{\rm COSMOS}$ (COSMOS bolometric AGN luminosity) 
we have two input parameters for Shen 2009 SMBH growth model. 
Evolution of AGN luminosities follows a universal general form 
of light curve with an initial exponential growth (pre-peak 
accretion) at constant Eddington ratio $\lambda$ = 3 until it 
reaches L$_{\rm peak}$, followed by a power-law decay. 
We replace L$_{\rm peak}$ with L$_{\rm COSMOS}$.

Note that there are two sets of 40,000 M$_{\rm BH,initial}$ 
and L$_{\rm COSMOS}$ for each $\Delta\rm z$-$\Delta\rm M_{*}$ 
interval, obtained from Monte Carlo realisations in two 
models: M1 (lower range of seed BH masses) and M2 (upper 
range of seed BH masses).

After applying best fit parameters of Shen 2009 to 
their equation 29, predicted SMBH mass (M$_{\rm BH, predicted}$), 
after AGN phase, can be written as:

\begin{equation}
{\rm M_{BH, predicted}} = \frac{{\rm L_{\rm COSMOS}}}{3l_{\rm Edd}} (1-\frac{2{\rm lnf}}{3})  , 
\end{equation}
and
\begin{equation}
f = \frac{3 l_{\rm Edd} {\rm M_{BH, initial}}}{{\rm L_{COSMOS}}}
\end{equation}
where $l_{\rm Edd}$=1.26 $\times$ 10$^{38}{\rm ergs}^{-1}\Msun^{-1}$.

\begin{figure*}
\vspace{10mm}
\epsfig{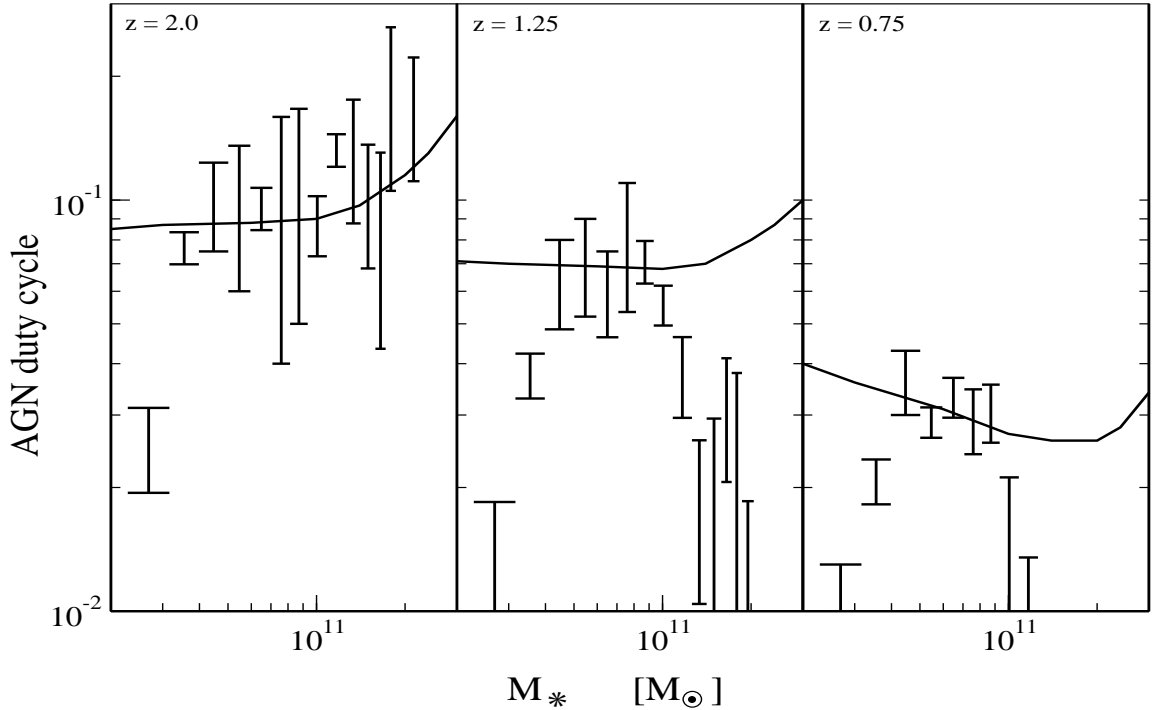}
\caption{AGN duty cycle as a function of stellar mass at three redshifts 
z = [2.00, 1.25, 0.75]. We consider AGNs with log L$_{\rm X}$ [erg/s] $\geq$ 43, 
where X = [2 - 10] KeV. Horizontal and vertical bars show the full range 
for duty cycle in our Monte Carlo realisations. Overplotted as thick black 
line is duty cycle for the same luminosity range from observations 
(HELLAS2XMM) of La Franca et al. 2005 (presented in Fiore et al. 2012).
Our best fit model is a good match to the observations although we 
slightly underpredict duty cycle toward lower redshifts.}
\end{figure*}

Through Monte Carlo realisations we take into account: 
all possible seed values that could be assigned to the 
merging DMHs; all possible luminosities in each luminosity 
bin of Bongiorno et al. 2012; scatter in Ferrarese 2002 
M$_{\rm BH}$ - M$_{\rm DMH}$ relation; scatter in Shen 2009 
M$_{\rm BH, final}$ - M$_{\rm DMH, post}$ relation; and 
scatter in Rodriguez-Puebla et al. 2015 M$_*$ - M$_{\rm DMH}$ 
relation. In the last mentioned scatter, same halo can host 
a galaxy below or above log(M$_{*}$/$\Msun$) = 10.4. As 
the result, depending on the random draw from the scatter 
in each Monte Carlo realisation, some halos might drop 
from the analysis while others might join.  

At the end we have M$_{\rm BH,final}$ from our best fit 
model and in 40,000 Monte Carlo realisations we produce 
M$_{\rm BH,predicted}$ in each $\Delta\rm z$-$\Delta\rm M_{*}$ 
interval. Now we can compare these two masses. If the observed 
AGN luminosities are indeed the peak luminosities when most of 
the SMBH growth occurs, then the mass of the predicted SMBH 
should match the mass of the final SMBH. We calculate the 
percentage of realisations when this condition is met.

\section{RESULTS}

\subsection{Best fit model}

We apply Shen 2009 major merger driven AGN activity model
to the merger trees in cosmological N-body simulation. 
There are some differences between Shen 2009 and our model. 

While Shen 2009 assumes constant ratio of 10$^{-3}$ between 
initial and peak BH mass, we seed DMHs with BH seeds and 
follow their evolution before AGN phase. Hence, BH mass 
right before AGN phase is not necessarily a constant 
fraction of the peak BH mass. Also, we calibrate final SMBH 
mass at redshift z = 0 to the local Ferrarese relation. In 
this manner, SMBH mass is overestimated at high redshift 
to accommodate for the late DMH evolution. In our model, super-Eddington
accretion starts when galaxies merge while in Shen 2009 model
same occurs when DMHs merge.

Our best fit model for the SMBH growth reproduces observed
AGN luminosity function, SMBH mass function, duty cycle, and
bias. Both M1 and M2 models can be considered as best fit models. 
M2 model provides a slightly better fit to the observations 
hence we show this match for M2 model only.

\begin{figure*}
\vspace{10mm}
\epsfig{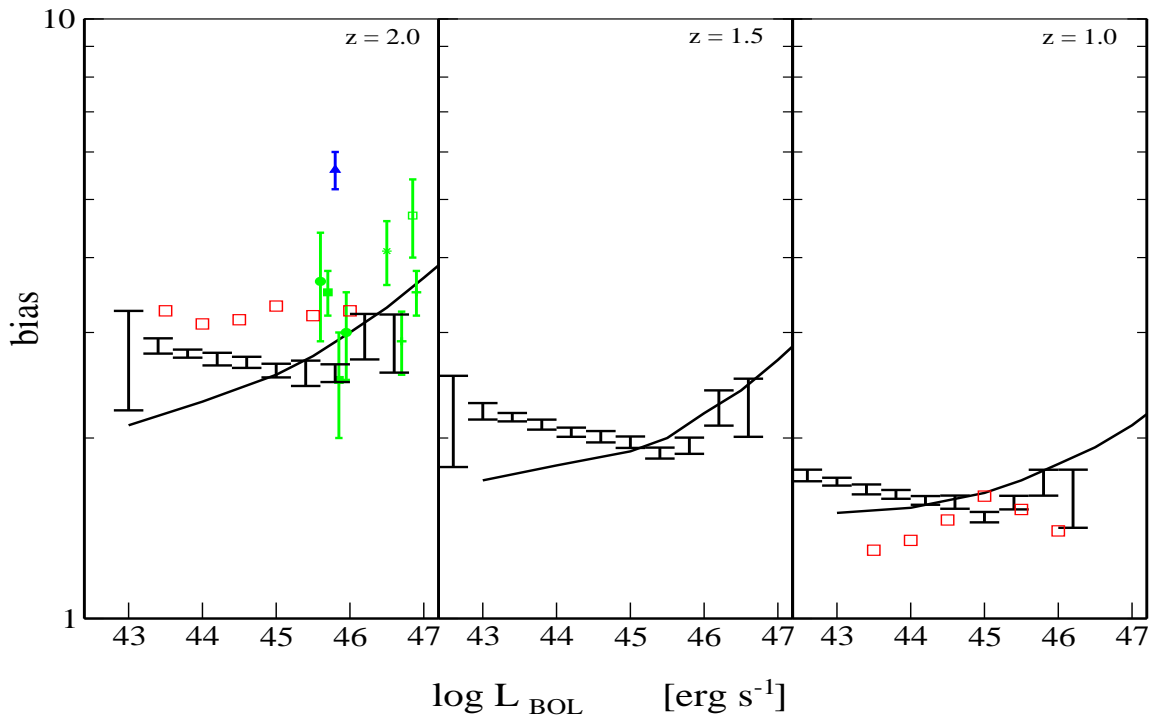}
\caption{AGN bias factor at three redshifts z = [2.0, 1.5, 1.0]. 
Horizontal and vertical bars show the full range for AGN bias 
in our Monte Carlo realisations. Thick black line shows the best fit model
in Shen 2009. Points are measurements from Croom et al. (2005, green-crosses), 
Porciani $\&$ Norberg (2006, green-star), Shen et al. (2009, green-open square), 
da Angela et al. (2008, green-circles), Myers et al. (2007, green-squares), 
and Allevato et al. 2011 (blue triangle). Red squares are from the semi
analytic model of galaxy formation in Gatti et al. 2016.
Considering the uncertainties in determining AGN bias, our best fit model 
is a good match to the observations.}
\end{figure*}

Figure 2 shows BH mass function at three redshifts 
z = [2.00, 1.25, 0.75]. Horizontal and vertical bars show
the full range for BH mass function, in our Monte Carlo realisations, for 
active black holes only, in AGNs with log L$_{\rm X}$ [erg/s] $\geq$ 43, 
where X = [2 - 10] KeV. 
Dotted, blue line shows our BHs mass function for all BHs. 
Overplotted as thick black line is active BH mass function 
for the same luminosity range from observations (HELLAS2XMM)
of La Franca et al. 2005 (presented in Fiore et al. 2012). 
Our best fit model follows the observed mass functions for 
active BHs. We slightly overestimate masses of active BHs 
at z = 2. This effect transfers to the lower redshift where our
BH mass function for all BHs (dotted, blue) slightly overpredicts
the local BH mass function at $\sim$ 10$^8\Msun$ (dashed, red, 
Merloni $\&$ Heinz 2008). At larger BH masses our
model underpredicts local BH mass function. We find that this 
occurs due to the arbitrary cut off at the higher mass end for
halos capable of hosting AGNs. When this upper limit for halo mass
is doubled, we get a perfect match to the local BH mass function
for M$_{\rm BH}$ $>$ 10$^8\Msun$. However, at the same time, we 
overpredict AGN luminosity function at z$<$1 and 
log L$_{\rm X} \rm [erg/s]$ $>$ 44 by a factor of 4.
Similarly to Shen 2009, our model is 
incomplete at M$_{\rm BH}$ $\leq$ 10$^{7.5} \Msun$ because we 
did not include contributions from AGNs triggered by secular 
processes or minor mergers.

Figure 3 shows AGN luminosity function with horizontal and 
vertical bars presenting the full range in our best fit model. 
Overplotted as thick black line is AGN luminosity function 
from same observations as in figure 2. Our best fit model 
deviates from the observations at z = 2 and z = 0.75. However, 
AGN luminosity functions reported in the literature deviate 
between various surveys. This can be seen when we overplot 
AGN luminosity function (dashed red line in Figure 3) from a 
large combination of X-ray surveys including XMM and Chandra
COSMOS survey (Miyaji et al. 2015). Discrepancy between Fiore 
et al. 2012 and Miyaji et al. 2015 is comparable to the 
discrepancy between our best fit model and these observations.

Figure 4 shows AGN duty cycle as a function of stellar 
mass at three redshifts z = [2.00, 1.25, 0.75]. We consider 
AGNs with log L$_{\rm X}$ [erg/s] $\geq$ 43, where X = [2 - 10] KeV. 
Horizontal and vertical bars show the full range for the duty 
cycle in our Monte Carlo realisations. Observations are again 
from La Franca et al. 2005 and Fiore et al. 2012. Our best 
fit model is a good match to the observations.

Figure 5 shows AGN bias factor at three redshifts 
z = [2.0, 1.5, 1.0]. We have calculated AGN bias factor
by using equations (3), (4), and (5) in Cappelluti, Allevato
$\&$ Finoguenov 2012. From these equations, AGN bias in a luminosity and redshift
range $\Delta \rm L, \Delta z$ can be
written as:

\begin{equation}
\rm bias(\Delta \rm L, \Delta z) = \frac{\Sigma \rm b_{\rm DMH}(\Delta \rm L, \Delta z)}{N_{\rm AGN}(\Delta \rm L, \Delta z)} , 
\end{equation}
where b$_{\rm DMH}$($\Delta \rm L, \Delta z$) is the large scale bias of dark matter halos
which host AGNs in the luminosity and redshift range $\Delta \rm L, \Delta z$, and N$_{\rm AGN}$($\Delta \rm L,\Delta z$)
is the total number of AGNs hosted by DMHs in the luminosity and redshift range $\Delta \rm L, \Delta z$.
We obtain b$_{\rm DMH}$ from figure 11 in Allevato et al. 2011.

\begin{figure}
\vspace{10mm}
\includegraphics [height=3.0in,width=3.0in,angle=0]{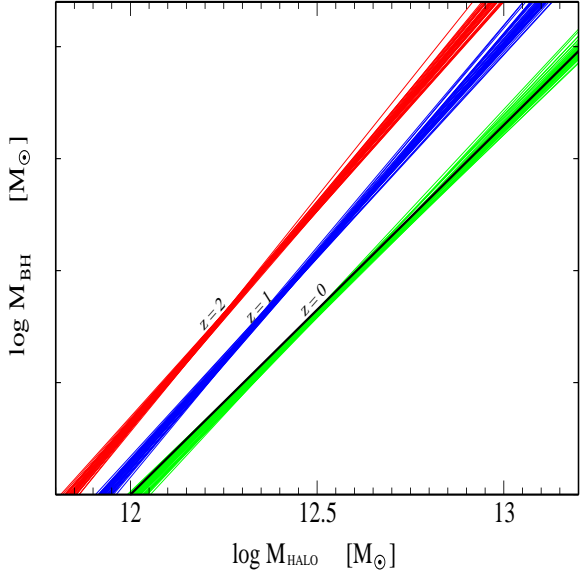}
\caption{M$_{\rm BH}$ - M$_{\rm DMH}$ relation in our best 
fit model. Red lines show full range of Monte Carlo 
realisations at redshift z = 2; blue lines represent z = 1; 
and green lines z = 0. Thick black line shows local Ferrarese 
relation at z = 0. Figure shows how M$_{\rm BH}$ - M$_{\rm DMH}$ 
relation evolves into local Ferrarese relation as dark matter 
halos grow in mass.}
\end{figure}

Horizontal and vertical bars in figure 5 show 
the full range for AGN bias in our Monte Carlo realisations. 
Thick black line shows the best fit model in Shen 2009. 
Points are measurements from Croom et al. (2005, green-crosses), 
Porciani $\&$ Norberg (2006, green-star), Shen et al. 
(2009, green-open square), da Angela et al. (2008, green-circles), 
Myers et al. (2007, green-squares), and Allevato et al. 2011 
(blue triangle). Red squares are from the semi analytic model 
of galaxy formation in Gatti et al. 2016. AGN bias
substantially differs from bias in Shen 2009 at low luminosities. 
This flattening could be induced by the Monte Carlo approach
as it includes the broad scatters in calculated parameters.
Considering the uncertainties in determining AGN bias, our best fit model 
is a good match to the observations.

\begin{figure}
\vspace{10mm}
\includegraphics [height=3.0in,width=3.0in,angle=0]{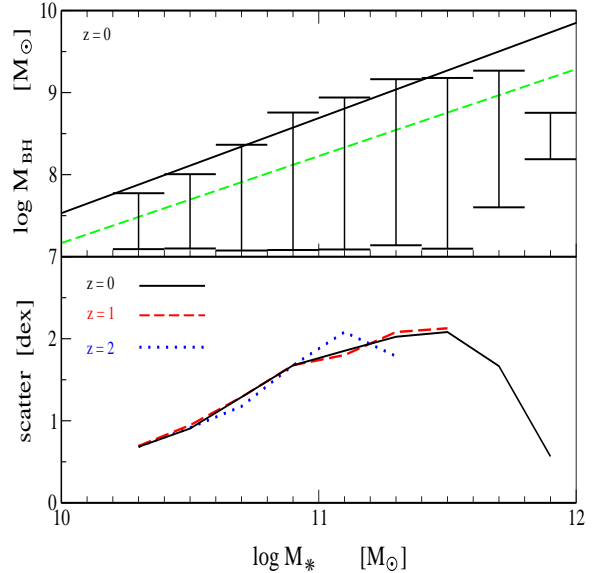}
\caption{Top panel: M$_{\rm BH}$ - M$_*$ relation in our best fit model.
Horizontal and vertical bars show full range of Monte Carlo realisations 
at redshift z = 0. Black line shows Kormendy $\&$ Ho 2013 relation,
and dashed green line shows Merloni $\&$ Heinz 2008 relation. 
Bottom panel: evolution of scatter in M$_{\rm BH}$ - M$_*$ relation in 
our best fit model (dotted blue line for z = 2; dashed red line for z = 1;
and thick black line for z=0). }
\end{figure}

Figure 6 shows M$_{\rm BH}$ - M$_{\rm DMH}$ relation in our 
best fit model. Red lines show full range of Monte Carlo 
realisations at redshift z = 2; blue lines represent the 
same at z = 1; and green lines at z = 0. Thick black line
shows local Ferrarese relation at z = 0 and it matches our 
best fit model at z = 0 by the default since we calibrate 
our model to do exactly that. We find that this match occurs 
when $\sigma_{\rm sph}$ = 0.77 $\times$ V$_{\rm vir}$. Our 
model incorporates no evolution in M$_{\rm BH}$ - $\sigma_{\rm sph}$ 
relation. It overpredicts BH mass at high redshift as BHs 
grow faster than DMHs. Figure 6 shows how M$_{\rm BH}$ - M$_{\rm DMH}$ 
relation evolves into local Ferrarese relation as dark matter 
halos grow in mass and ``catch up'' to the BH growth.

Figure 7 shows M$_{\rm BH}$ - M$_*$ relation in our best 
fit model (top panel). Horizontal and vertical bars show 
full range of Monte Carlo realisations at redshift z = 0. 
Black line shows Kormendy $\&$ Ho 2013 relation, and dashed 
green line shows Merloni $\&$ Heinz 2008 relation. Our best 
fit model underpredicts BH masses when compared to Kormendy 
$\&$ Ho 2013 relation. However, the match is better when 
compared to Merloni $\&$ Heinz 2008 relation. We also find 
no evolution of scatter in M$_{\rm BH}$ - M$_*$ relation in 
our best fit model (bottom panel in figure 7).
Despite the BH mass being determined by $\sigma_{\rm sph}$ via V$_{\rm vir}$,
and the scatter in M$_*$ at fixed M$_{\rm DMH}$ is very small, the resulting 
M$_{\rm BH}$ - M$_*$ relation of figure 7 is very broad and even 
significantly below the Kormendy $\&$ Ho 2013 relation. 
This might be in support of the biases in the local scaling relations of 
BHs and galaxies discussed recently in the literature 
(Reines $\&$ Volonteri 2015, van den Bosch et al. 2015,
Shankar et al. 2016, Greene et al. 2016, van den Bosch 2016).

\subsection{Determining AGN activity phase}

Assuming that major mergers are driving AGN activity in 
massive galaxies, we have selected simulated mergers of 
field galaxies in the redshift range 0.3 $<$ z $<$ 2.5 
and matched them to the observed samples of AGNs in 
redshift and galaxy-mass bins in F14.  

Matching procedure briefly consists of: As halo merger 
finishes, galaxy merger starts. We define that as a time 
of AGN peak activity corresponding to the AGN observed 
in COSMOS survey. The mass of simulated galaxy hosting 
the AGN is derived from halo-galaxy scaling relation. 

Once we find redshift bin and galaxy-mass bin of the 
simulated merger, we trace the merging halos before the 
merger, and we trace merger remnant after the merger. 
We determine initial SMBH mass (before accretion during 
AGN phase) and final (``true'') SMBH mass (after accretion 
in AGN phase).

Figure 8 shows BHs mass function in model M1 (left panels) 
and M2 (right panels) for both initial BHs M$_{\rm BH, initial}$ 
(thick, black line) and for the final BHs M$_{\rm BH, final}$ 
(thin, red line), in the three redshift ranges: top panel: 
0.3 $<$ z $<$ 0.8; middle panel: 0.8 $<$ z $<$ 1.5; bottom panel: 
1.5 $<$ z $<$ 2.5. Bars represent full range of Monte Carlo 
realisations which cover possible M$_{\rm BH, initial}$ from 
log-normal distribution defined for massive seed BHs in spiral 
galaxies; scatter in Ferrarese 2002 M$_{\rm BH}$ - M$_{\rm DMH}$ 
relation for BHs at the centres of elliptical galaxies; and 
scatter in Kormendy $\&$ Ho 2013 M$_{\rm BH}$ - $\sigma_{\rm sph}$ 
relation for the final ``true'' BH mass in postmerger halos. 
It also covers (not that obvious) scatter in Rodriguez-Puebla 
et al. 2015 M$_*$ - M$_{\rm DMH}$ relation applied as a selection 
criterion for elliptical galaxies at z$_{\rm AGN}$. Because of 
this scatter, same halo can host a galaxy below or above 
log(M$_{*}$/$\Msun$) = 10.4. As the result, depending on the 
random draw from the scatter in each Monte Carlo realisation, 
some halos might drop from the analysis while others 
might join. 

Premerger accretion occurs between z$_{\rm initial}$ and 
z$_{\rm AGN}$ and it consists of two phases: first, the 
pre-BHB phase before initial BHs form binary; and second, 
BHB phase which lasts for $\sim$ 100 Myr before BHs merge as 
BHs in the binary overcome last couple of kiloparsecs. 
The typical pre-BHB timescale for BHs in spiral galaxies is 
$\sim$ 2 Gyr. Since accretion rate in model M1 is set to 
10$^{-4}\Msun {\rm yr^{-1}}$ the amount of mass accreted 
then during this phase is $\sim$ few $\times$ 10$^5\Msun$. 
During BHB phase BHs double their masses. After adding mass 
from both pre-BHB and BHB phases to the seed BHs in spiral 
galaxies their mass function peaks at 10$^{5.5}\Msun$ - 10$^{6}\Msun$ 
depending on the redshift (figure 8, left panels). On the 
other hand, mass function of the initial BHs in the 
elliptic galaxies peaks at 10$^{6.5}\Msun$ - 10$^{7}\Msun$ 
(figure 8, left panels).

In model M2, the pre-BHB accretion is set to 
10$^{-3}\Msun {\rm yr^{-1}}$ so the amount of mass accreted 
during this phase is $\sim$ few $\times$ 10$^6\Msun$.     
After both pre-BHB and BHB phases, and after adding the 
accreted mass to the seed BHs in spiral galaxies, 
their masses overlap with the masses of BHs in elliptical 
galaxies. Resulting mass function peaks at 10$^{6.5}\Msun$ 
- 10$^{6.9}\Msun$ depending on the redshift (figure 8, right panels).

The difference in mass function between initial and final 
BHs in figure 8, is the accreted mass during AGN phase in 
our best fit model where L$_{\rm peak}$ is calculated
from the light curve model in Shen 2009.
M$_{\rm final}$ obtained in this manner (the ``true'' final 
BH mass) is then compared to M$_{\rm predicted}$ which is 
obtained by replacing L$_{\rm peak}$ with AGN 
luminosities from COSMOS survey L$_{\rm COSMOS}$.

All galaxies in the specific mass range, host AGNs with 
the probability defined in F14. Probability functions 
presented in F14 show that galaxies are more likely to 
host less luminous AGNs. As the observed AGN luminosity 
increases, the probability for that particular AGN to be 
observed in the COSMOS galaxy decreases. We incorporate 
COSMOS AGN luminosities into Shen 2009 model for SMBH growth.

We perform 40,000 Monte Carlo realisations for every 
M$_{\rm initial}$ in each $\Delta\rm z$-$\Delta\rm M_{*}$ 
interval, through all possible COSMOS AGN luminosities. As 
the result, we obtain 40,000 predicted BH masses which we 
compare to the ``true'' final BH mass. When 
M$_{\rm predicted}$ $\geq$ M$_{\rm final}$, COSMOS AGN 
luminosity is the peak AGN luminosity, corresponding to the 
AGN luminosity in our best fit model. We calculate 
the percentage of realisations where M$_{\rm predicted}$ 
is at least as large as M$_{\rm final}$ and present it in figure 9.

\begin{figure*}
\vspace{10mm}
\epsfig{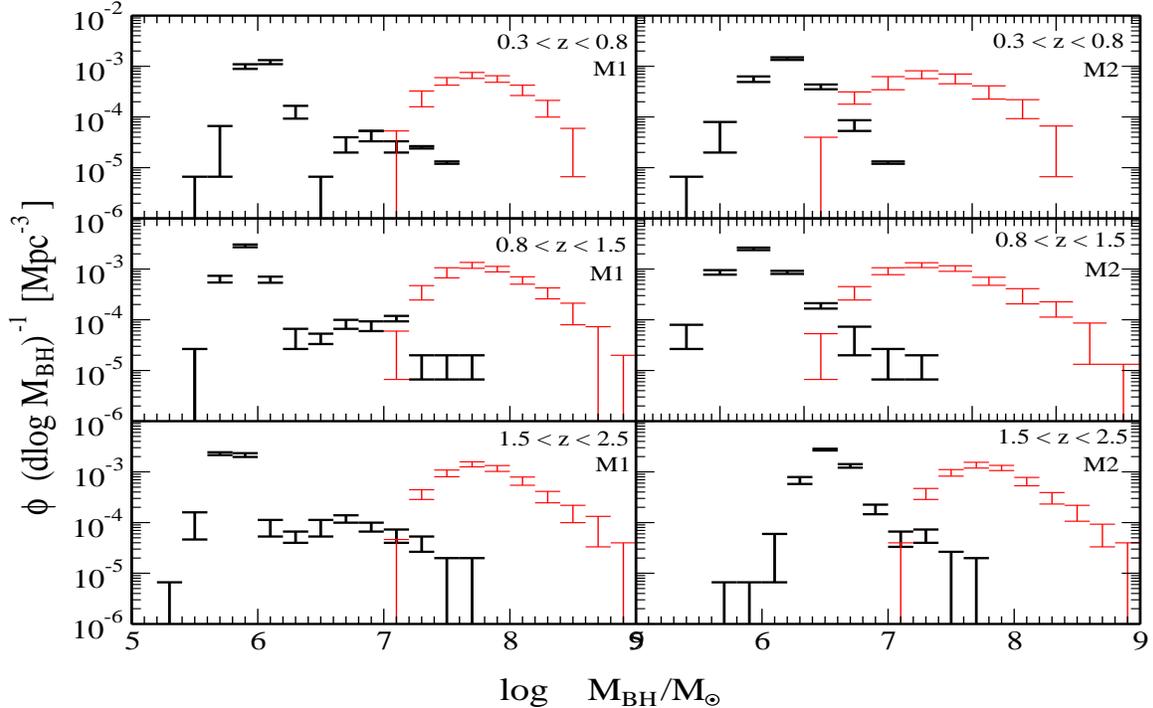}
\caption{BH mass function for model M1 with bars representing
full range of Monte Carlo realisations. M$_{\rm BH, initial}$ 
represented in thick, black. M$_{\rm BH, final}$ represented in 
thin, red. Three redshift ranges: top panel: 0.3 $<$ z $<$ 0.8; 
middle panel: 0.8 $<$ z $<$ 1.5; bottom panel: 1.5 $<$ z $<$ 2.5. 
BH mass function for model M2.}
\end{figure*}

Figure 9 shows the probability function (occupation fraction)
that the observed AGNs are at their peak activity. Nine panels 
present three redshift ranges and three galaxy log-mass ranges. 
Thick (black) bars represent probability functions in our model M1. 
Thin (red) bars represent probability functions in our model 
M2. Bars show full range of Monte Carlo realisations. 
Probability for peak AGN activity at low redshift z=[0.3, 0.8] 
is small in all galaxy mass bins. For M$_*$ = [10.4, 10.7] 
all AGNs have probability $<$ 20 $\%$ in model M1, and $<$ 30 $\%$ 
in model M2. In the same $\Delta\rm z$-$\Delta\rm M_{*}$ interval,
probability of $<$ 10 $\%$ have 90 - 100 $\%$ of AGNs in M1 and 
80 - 90 $\%$ of AGNs in M2. The occupation fraction of AGNs with 
low probability for peak activity increases toward larger galaxy 
mass. AGNs hosted by most massive galaxies (M$_*$ = [10.9, 11.2]) 
are all in the declining phase of their activity since probability 
drops to $<$ 20 $\%$ in both models (top, right panel in figure 9).  
Overall, at low redshift, almost all AGNs are in non-star-forming 
Red Sequence galaxies.

Increase in fraction of AGNs with larger probability means 
more AGNs are in star forming Green Valley galaxies. We see 
this trend as we go from low to high redshift in figure 9. 
At the intermediate redshifts (z=[0.8, 1.5]) AGN fraction with 
larger probability increases in both models (middle panels in 
Figure 9). As expected, this increase is larger for M2 where 
M$_{\rm BH, initial}$ is larger. Still, most AGNs have low 
probability for being observed at their peak. AGNs at high 
redshifts (z=[1.5, 2.5]), and in the lowest galaxy mass range 
(M$_*$ = [10.4, 10.7], bottom, left panel in figure 9) 
are dominantly at the peak activity since 30 - 50 $\%$ of them 
in M1 and 55 - 75 $\%$ in M2 have $>$ 80 $\%$ probability for 
being at the peak. Overall, distribution of occupation fractions 
shifts toward larger probabilities. Similarly to lower redshifts, 
occupation fraction with large probabilities decreases toward 
more massive galaxies. For M$_*$ = [10.7, 10.9] (bottom, 
middle panel in figure 9), AGN fraction is evenly distributed. 
Here we would expect to see comparable numbers of AGNs in both 
quiescent and star forming galaxies. In the largest mass range 
(panels on the right of Figure 9) AGNs are predominantly in 
quiescent galaxies at all redshifts.

So the trend that emerges in figure 9 is that quiescent 
galaxies host almost all AGNs at low redshift. As we go 
toward higher redshift there are more AGNs in star forming 
galaxies and the percentages of AGNs inhabiting quiescent 
galaxies and star forming galaxies become comparable. We 
also see the trend with increasing galaxy mass. At larger 
galaxy masses there are more AGNs in quiescent galaxies. 

This exact trend we see in AGNs in COSMOS survey 
(Figures 12, 13 and 18, Bongiorno et al. 2012). 

\begin{figure*}
\vspace{10mm}
\epsfig{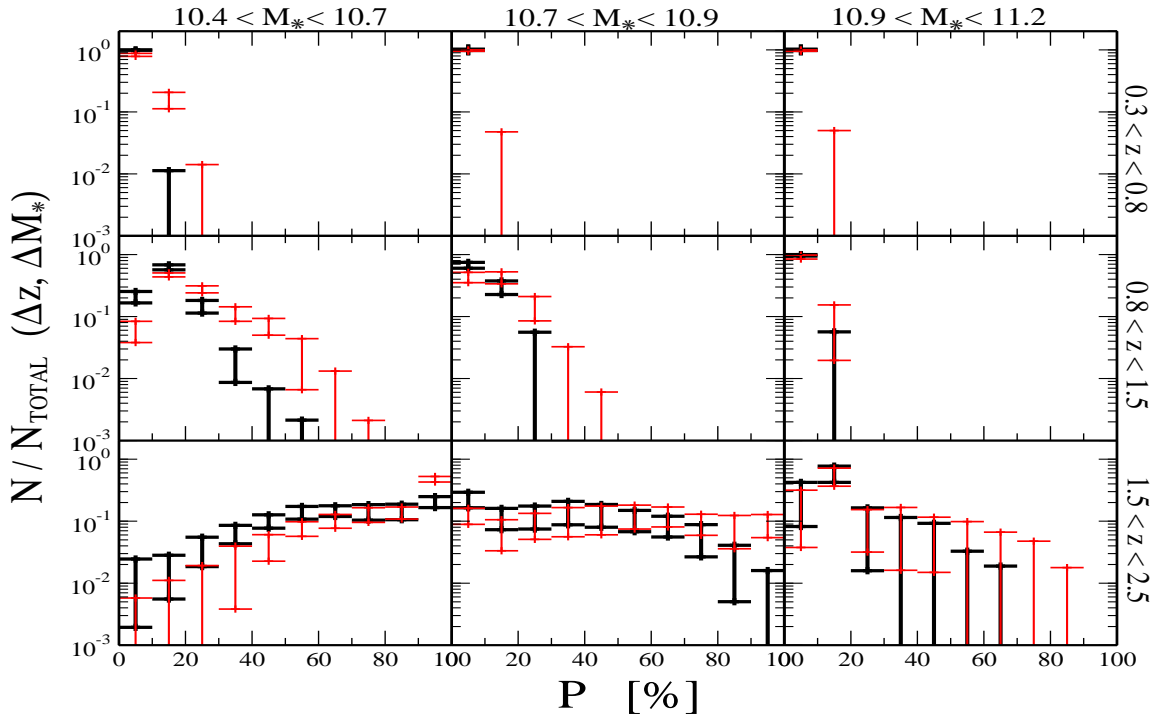}
\caption{Probability function for the predicted SMBH mass 
to be at least as large as the true SMBH mass. In other words, 
probability that the observed AGN luminosity is large enough to 
account for the final SMBH mass. Probability functions are split 
into redshift bins and galaxy mass bins which correspond to 
nomenclature in Bongiorno et al. 2012. Thick (black) histograms
represent probability functions in our model M1. Thin (red) 
histograms represent probability functions in our model M2.}
\end{figure*}

\section{DISCUSSION AND CONCLUSIONS}

We ran cosmological (130 Mpc box) N-body (dark matter only) 
simulation from which we located field DMHs at all redshifts. 
We also followed their evolution while they stay in the field. 
We found merger events and traced merger progenitors and 
merger remnants. Through scaling relations we calculated SMBH 
masses for progenitors and remnants. In this manner we obtain 
the SMBH mass at the centres of DMHs before (initial SMBH) 
and after (final SMBH) the merger.

We assume that at the time when halo merger finishes, galaxy 
merger starts. At that time newly formed SMBH ignites as AGN 
and quickly reaches its peak activity. We focus on two models 
with different range for the initial BH mass since BH seeds 
in spiral galaxies and there pre-coalescence growth are the 
source of largest uncertainty in our modelling. Model M1 has 
a lower mass range $\sim$ [10$^5$ - 10$^6$] $\Msun$ and 
pre-coalescence accretion rate of 10$^{-4} \Msun {\rm yr^{-1}}$. 
Model M2 has a larger initial mass range $\sim$ [10$^{5.5}$ - 
10$^{6.5}$] $\Msun$ and accretion rate of 10$^{-3} \Msun {\rm yr^{-1}}$. 

We determine ``true'' final BH mass by using non-evolving 
M$_{\rm BH}$ - $\sigma_{\rm sph}$ relation where 
$\sigma_{\rm sph}$ = 0.77 $\times$ V$_{\rm vir}$. 
In this manner, M$_{\rm DMH}$ - M$_{\rm BH}$ relation evolves 
from overestimating BHs masses at high redshift to matching
local Ferrarese relation at z=0.

Our best fit model for the SMBH growth reproduces observed
AGN luminosity function, SMBH mass function, duty cycle and
bias.

Next, we replace peak AGN luminosities in our best fit 
model with COSMOS AGN luminosities from Bongiorno et al. 2012.

For every galaxy hosting an AGN we determine redshift and 
mass, and sort them into redshift ranges and mass ranges as 
in Bongiorno et al. 2012, COSMOS survey. For each mass and 
redshift, we assign an observed probability function for a 
galaxy to host an AGN of a certain luminosity (Figure 14 
in Bongiorno et al. 2012). Next we ran 40,000 Monte Carlo 
realisations in each $\Delta\rm z$-$\Delta\rm M_{*}$ 
interval where we draw from the observed probability 
functions and we assign luminosities to the initial BH. 
We obtain 40,000 predicted BH masses which we compare to 
the ``true'' final BH mass.

When M$_{\rm predicted}$ $\geq$ M$_{\rm final}$, COSMOS 
AGN luminosity is the peak AGN luminosity, corresponding 
to the peak AGN luminosity in our best fit model. We 
calculate the percentage of realisations where M$_{\rm predicted}$ 
is at least as large as M$_{\rm final}$. Large percentage 
implies large probability for AGNs to be at their peak 
activity. Small percentage means that AGNs are not observed 
at the peak but in the declining phase of their nuclear 
activity. In this manner, we distinguish ``peak'' AGNs 
(recently merger triggered and hosted by star forming 
galaxies, Green Valley) and ``faded'' AGNs (merger triggered 
a long time ago and now residing in quiescent galaxies, 
Red Sequence).

At low redshift range (z=[0.3, 0.8]) all observed AGNs 
are in the declining phase of their nuclear activity, 
fading away (figure 9). The probability for being at 
their peak activity is $<$ 10 $\%$ for $>$ 90 $\%$ of 
AGNs in the most massive galaxies. AGN luminosity would 
have to be very large to account for the SMBH growth.
But even if this highest possible luminosity was large 
enough to produce final SMBH, it is also the least 
probable one. Since the entire range of luminosities 
can not produce final SMBH, then these luminosities do 
not correspond to the AGN peak activity. The time of 
maximum nuclear activity when most of the mass was 
accreted has occurred in the past at higher Eddington 
ratio when AGN luminosity was larger and when AGN was 
most likely hosted by star-forming galaxy (Green Valley). 
Logical conclusion is that the observed luminosities 
belong to the AGN in the declining phase of its nuclear 
activity which places this particular AGN in the 
quiescent galaxy (Red Sequence).

Theoretical modelling of AGN populations in hosts of 
various morphologies, mass ranges and redshifts, support 
the merger driven scenario for luminous AGN activity. At 
the same time observations are split between existence 
of merger features (Schawinski et al. 2010, Smirnova, 
Moiseev $\&$ Afanasiev 2010, Koss et al. 2010, Cotini et 
al. 2013) and the lack of them (Gabor et al. 2009, Darg et 
al. 2010, Cisternas et al. 2011, Kocevski et al. 2012, 
Villforth et al. 2014). Villforth et al. 2014 found no 
increase in the prevalence of merger signatures with AGN 
luminosity (in the redshift range z = [0.5, 0.8]) and 
concluded that major mergers either play only a very 
minor role in the triggering of AGN in the luminosity 
range studied (log L$_{\rm X}$ = [41, 44.5]) or time 
delays are too long for merger features to remain visible.

Our model shows that the merger driven scenario is still 
consistent with the observations even though there are no 
merger features in massive galaxies hosting low redshift 
AGNs and almost all of the AGN hosts are quiescent galaxies. 
How can mergers explain AGN activity in massive galaxies 
which have no merger features and no star formation to 
indicate recent galaxy merger? Since at z = [0.3, 0.8] 
(figure 9) the observed luminosities can not correspond to 
AGNs at their peak activity (can not produce final SMBH 
mass in the simulation), then they must be observed much 
later in their evolution long after the merger features can 
be detected. And our confirmation of Bongiorno et al. 2012 
results that almost all low redshift AGNs are in quiescent 
galaxies is a simple consequence of the drop in galaxy 
merger rates at z $<$ 1. Since galaxy merger rates fall 
dramatically at low redshift, there are very few recently 
activated AGNs which would be hosted by star forming 
galaxies. So most of the observed AGNs are the fading 
AGNs activated in the old mergers which occurred at higher 
redshifts. Since there are no new galaxy mergers, almost 
all observed AGNs are in non-star-forming galaxies.

As we go toward higher redshifts, the probability for the 
AGNs being at their peak activity increases. There are more 
AGNs in star-forming Green Valley galaxies. At z=[1.5, 2.5] 
the percentage of AGNs in star forming galaxies is 
comparable to the percentage of AGNs in quiescent galaxies. 
This can be seen in our figure 9, bottom panels, and in 
Figure 18 of Bongiorno et al. 2012. Again, this is a simple 
consequence of the large merger rate in galaxies at high 
redshift. The explanation for comparable number of star 
forming and quiescent AGN hosts is that AGNs in star forming 
galaxies at high redshift have ``just'' been triggered by 
galaxy mergers while AGNs in quiescent galaxies at the same 
redshift have been merger triggered at some time in the past.

Schawinski et al. 2014 had proposed a split of Green Valley 
transition into two paths. Current understanding is that 
late type galaxies transition slowly from Blue Cloud to Red 
Sequence, while hosting low to intermediate luminosity AGNs 
driven by secular processes. Early type galaxies transition 
fast, while hosting high luminosity AGNs driven by major mergers. 
In the context of galaxy evolution, our model addresses the 
evolution of early type galaxies which are produced in major 
mergers of gas-rich disk/spirals. These galaxies correspond 
to the massive, red galaxies in COSMOS survey (Bongiorno et 
al. 2012) where they represent the majority of AGN hosting 
galaxies. According to our model, AGNs in massive galaxies 
of the COSMOS survey, belong to the rapid transition channel 
(Schawinski et al. 2014). We find that, right after the merger, 
AGNs reach their peak activity (Green Valley phase). This is 
a short phase ($\sim$ 100 Myr) during which star formation is 
quenched. Then, galaxies enter Red Sequence phase with AGNs 
in the decline (or at the end) of their nuclear activity and 
low Eddington accretion rate observed in COSMOS survey. 
Figure 2 shows that we are sampling the growth of SMBHs 
$>$ 10$^7 \Msun$. For the most part, final SMBHs are $>$ 10$^8 \Msun$. 
This is consistent with Hopkins, Kocevski $\&$ Bundy 2014 
conclusion that at these masses merger driven AGN activity dominates.

There are a number of recent papers discussing possible
biases in the local scaling relations of BHs and galaxies
(Reines $\&$ Volonteri 2015, van den Bosch et al. 2015,
Shankar et al. 2016, Greene et al. 2016, van den Bosch 2016).
In support of this, we find that when $\sigma_{\rm sph}$ 
is determined via V$_{\rm vir}$, the resulting 
M$_{\rm BH}$ - M$_*$ relation of figure 7 is very broad and even 
significantly below the Kormendy $\&$ Ho 2013 relation. 
Shankar et al. 2016 found that the normalisation of the 
M$_{\rm BH}$ - $\sigma$ relation might be decreased by a factor 
of 3. Revising M$_{\rm BH}$ - $\sigma$ relation in our best fit 
model leads to smaller final BH masses. This in turn
decreases BH mass functions and AGN luminosity functions by a similar
factor but still consistent with the observations. 
We have tested how this fact would influence our results and we
found that the probability functions in figure 9 would shift toward
higher probabilities but would not qualitatively change our results.

We conclude that merger driven scenario for AGN activity is 
consistent with the observations and that the occupation 
fractions of the observed AGNs simply follow the evolution 
of galaxy merger rates. Our model reproduces the observed 
trend that quiescent (Red Sequence) galaxies host almost 
all AGNs at low redshift due to the dramatic drop in galaxy 
merger rates at z $<$ 1. There are just few recently 
activated AGNs in star forming galaxies. Instead, most AGNs 
are in their declining nuclear activity hosted by quiescent 
galaxies. As we go toward higher redshift (z $>$ 1), galaxy 
merger rates increase, and there are more peak activity AGNs 
observed in star forming galaxies. The percentage of peak 
AGNs inhabiting star forming and the percentage of faded AGNs 
hosted by quiescent galaxies becomes comparable. We also 
confirm the observed trend with increasing galaxy mass. At 
larger galaxy masses there are more AGNs in quiescent galaxies.

Our method for matching simulated DMH merger events with 
observations of field AGNs will be more accurate as the 
statistics improves with the future surveys. At this point, 
limited statistics of the sample prevents more detailed 
investigations of the incidence of AGN in galaxies as a 
function of redshift, stellar mass, star-formation rate 
and nuclear luminosity (Bongiorno et al. 2012). Wide area 
surveys will be necessary to probe volumes at z $>$ 1 
comparable to that explored by SDSS. At the low end of the 
galaxy mass distribution log [10.4, 10.7] $\Msun$, and high redshift 
[1.5, 2.5], the probability functions in Bongiorno et al. 
2012 do not have AGNs with log L$_{\rm X}$ [erg/s] $\leq$ 44. Most 
likely missed in COSMOS survey, since there are AGNs in this 
luminosity and galaxy mass range at lower redshifts. Including 
lower luminosities would change probability functions in 
Figure 14 of Bongiorno et al. 2012, and would most likely 
shift AGN occupation fractions toward lower probabilities.

Even though we enforce criterion on AGN hosts
to be $>$ 10$^{10.4} \Msun$ in stellar mass, we must have some
mixing of populations. We expect that a large majority
of AGNs in these galaxies are merger driven. However, some 
percentage of AGNs is probably driven by secular processes.
That being said, we would like to point out that we are not 
trying to show that mergers are definitely responsible for 
AGN activity. We are arguing that observed AGN activity 
(at least in $>$ 10$^{10.4} \Msun$ galaxies) is consistent with
mergers as drivers. However, this does not exclude other 
mechanisms. In fact, one could imagine a scenario where all 
of the moderate to faint AGNs are secularly driven.

Two major concerns for our method are: precision in 
determining the mass of the AGN host galaxy in 
observations, and detecting low luminosity AGNs. 
Both concerns impact the relations between mass of 
the host galaxy and probability functions for AGN 
incidence.

\section*{ACKNOWLEDGMENTS}

Authors would like to thank the anonymous reviewer 
for the tremendous help with shaping this paper.

This work was supported by the Ministry of Education, Science and 
Technological Development of the Republic of Serbia through project 
no. 176021, ``Visible and Invisible Matter in Nearby Galaxies: Theory 
and Observations''. The author acknowledges the financial support 
provided by the European Commission through project BELISSIMA 
(BELgrade Initiative for Space Science, Instrumentation and Modelling 
in Astrophysics, call FP7-REGPOT-2010-5, contract no. 256772).

Numerical results were obtained on the PARADOX cluster at the 
Scientific Computing Laboratory of the Institute of Physics Belgrade, 
supported in part by the national research project ON171017, funded 
by the Serbian Ministry of Education, Science and Technological Development.

\end{document}